\documentstyle[12pt]{article}
\begin{titlepage}   
\title{ 
{
\vspace{-8.5ex}
\begin{flushright}
\small\bf IHEP \normalsize \bf 2000--12\\
\end{flushright}
\vspace{10ex}
}
{\bf  Parametrizing and rephasing}\\
{\bf  neutrino mixing }\\[4ex]
}
\author{ Yu.~F.~Pirogov{}\thanks{E-mail: pirogov@mx.ihep.su}
\\[1ex]
{\it Institute for High Energy Physics,}\\
{\it Protvino, RU-142284 Moscow Region, Russia}\\[0.5ex]
{\it  Moscow  Institute  of  Physics  and  Technology,}\\
{\it  Dolgoprudny,  Moscow  Region, Russia }
}
\date{}

\begin{document}
\maketitle
\thispagestyle{empty}

\abstract{
\noindent
Neutrino mixing in the 
standard model extensions, both renormali\-zable and effective,
with arbitrary numbers of the singlet and left-handed
doublet neutrinos is investigated in a systematic fashion. 
The char\-ged and neutral (the $Z$ and Higgs mediated)  lepton
currents are written under general Majorana condition, and  the
odservable independence of the choice of the condition, the rephasing
invariance, is studied. A parame\-trization of the
neutrino  mixing matrices in
the doublet-singlet factorized form is developed. 
Its relationship with the see-saw mechanism is shown in the
limit of a small doublet-singlet mixing. The structure of the 
mixing matrices relevant to the neutrino  oscillation experiments is 
explicated.
}
\thispagestyle{empty}
\end{titlepage}

\def\l{\varphi}
\def\kappa{\omega}
\setcounter{page}{2}

\section{ Introduction}

The lepton sector of the minimal Standard Model (SM) of  electroweak
interactions is amazingly simple and symmetric.
Due to the absence of the right-handed neutrinos and neutrino
masses, the SM predicts  no flavour and $CP$
violation for leptons. Nevertheless, there is no known rule which
would prohibit
neutrinos from  acquiring masses. More than this, there are numerous
indications of the contrary. If so, the  lepton mixing has to  take
place  with all the subsequent phenomena such as flavour and $CP$
violation, neutrino oscillations, etc (as a recent review see, e.g.,
ref.~\cite{now}).

The  lepton mixing, unlike the quark one, should   generally be 
much more complicated. There are two main reasons for this. First,
the number of the (iso)singlet
neutrinos is a priori arbitrary relative to that of the
(iso)doublet ones. Second,  the Majorana masses for neutrinos are
possible in addition to the Dirac ones.  
As a result, three types of associated problems arise. First,
what is the total number of  physical parameters, and how  many of
them are masses,  mixing  angles and $CP$ violating phases? Second,
what do the lepton currents, both the vector and scalar
ones, look like in terms of the mixing  matrices? And third, how to
parametrize the  matrices explicitly?
In the previous paper~\cite{nu1} (see also references therein)  we
have systematically studied the parameter counting problem for the SM
extensions, both renormalizable and effective,  with
arbitrary numbers of the singlet and  left-handed doublet
neutrinos. Here we address ourselves to the second and  third
problems.

The gauge  interactions  of Majorana neutrinos
for the SM extensions with  arbitrary numbers of the singlet and
left-handed doublet neutrinos were studied in ref.~\cite{nm}, where  a
parametrization of the neutrino mixing matrices was also proposed. The
Yukawa neutrino interactions within the framework of the
renormalizable SM extensions with an equal number
of the singlet and doublet neutrinos were considered  in
ref.~\cite{pilaft}. The studies of refs.~\cite{nm,pilaft} were carried
out traditionally under  canonical Majorana condition. 
In the present paper these results are generalized under arbitrary
Majorana condition for any  SM extensions, both renormalizable
and effective, with arbitrary  numbers of the singlet and
left-handed doublet neutrinos. 
The freedom of the choice of the Majorana condition, the rephasing
invariance, is put as a corner-stone of  the whole study. 
Some  of our results  are  known in the
literature in one form or another. Nevertheless, having being
extended, they are hoped to be presented in the paper in a more
systematic fashion.

In Section~2, the structure of the neutrino interactions, both gauge
and Yukawa, are studied under arbitrary Majorana condition. In
Section~3, the  properties of the
interactions under  Majorana  neutrino rephasing, including
requirements for $CP$ invariance, are considered.
A parametrization of the mixing 
matrices in the doublet-singlet factorized form is proposed in
Section~4.  Its  relationship, under small doublet-singlet
mixing, with the see-saw mechanism~\cite{see-saw} is shown.
And finally, the patterns of  neutrino mixing matrices, relevant to
neutrino oscillation experiments, are discussed in Section~5.

\section{Lagrangians and mixing matrices}

\paragraph{(i) Weak  basis}

The most general renormalizable
$SU(2)_{\mbox{\scriptsize W}}\times
U(1)_{\mbox{\scriptsize Y}}$ invariant  lepton
Lagrangian of the SM extended by  the right-handed neutrinos  reads
\def\d{\partial\hspace{-1.2ex}/\hspace{0.3ex}}
\def\D{D\hspace{-1.5ex}/\hspace{0.5ex}}
\begin{eqnarray}\label{eq:lagrangian}
{\cal L}&=&
~~~ \overline{ l_L^{\tiny{0}}}i\D  l_L^{\tiny{0}} +  \overline{
e_R^{\tiny{0}}}i\D
e_R^{\tiny{0}} 
  +  \overline{ \nu_R^{\tiny{0}}}i\d  \nu_R^{\tiny{0}}\nonumber\\
&& - \Big{(} \overline{ l_L^{\tiny{0}}} Y^e  e_R^{\tiny{0}}\phi 
           + \overline{l_L^{\tiny{0}}} Y^\nu  \nu_R^{\tiny{0}} \phi^C
           + \frac{1}{2} \overline{ \nu_L^{\tiny{0}C}} {M^\dagger}
           \nu_R^{\tiny{0}}
           + \mbox{h.c.}
     \Big{)}\,. 
\end{eqnarray}
In eq.~(\ref{eq:lagrangian}), the   
lepton doublet $l_L^0$  and singlet $e_R^0$, $\nu_R^0$
fields with a zero superscript mean those in a weak basis
where, by definition, the symmetry properties are well stated.
It is supposed that  the ordinary chiral families of the SM
with the doublet left-handed   Weyl neutrinos in
number $d\ge 3$  are added by the singlet (sterile) Weyl
neutrinos in number $s\geq 0$.  
Let us designate such a renormalizable SM  extensions as
$(d,s)_{\mbox{\scriptsize r}}$. 
A priori, one should  retain $s$ and
$d$ as arbitrary integers, both  $s\leq d$ and  $s>d$ being
allowed.\footnote{We omit in the present analysis the possible
vector-like lepton doublets. Hence, with account for the most
probable exclusion
of the fourth heavy chiral family~\cite{4f}, one should put in
reality $d=3$. Nevertheless, we
retain $d$ as a free parameter to better elucidate the parameter
space structure of the extended SM.}
Further, $\D\equiv\gamma^\alpha D_\alpha$ is the generic covariant
derivative
which reduces to the ordinary one, $\d =
\gamma^\alpha\partial_\alpha$, 
for the hypercharge zero singlet neutrinos.
Here and in what follows the notations  ${\nu}^{\tiny{0}C}_L
\equiv
({\nu}_R^{\tiny{0}})^C  = C\overline{{\nu}_R^{\tiny{0}}}\,{}^T$, etc,
are used for the particle-antiparticle conjugates of chiral fermions
in the weak basis.
$Y^e$ and $Y^\nu$ are the arbitrary complex $d\times d$ and
$d\times s$ Yukawa  matrices, respectively, and 
$M$ is a complex symmetric $s\times s$ matrix of the Majorana masses
for the singlet neutrinos.
Finally,  $\phi$ is the Higgs isodoublet and $\phi^C\equiv i\tau_2
\phi^*$ is its charge conjugate.

One can generalize the preceding
considerations to  the most exhaustive
Dirac-Majorana case with the left-handed  Majorana masses. 
The direct Majorana mass term for the
doublet neutrinos is excluded in the minimal SM by the symmetry
and
renormalizability requirements. But  in the extended
SM as a low energy effective theory it could stem 
from the SM invariant operator  of the fifth dimension 
\begin{equation}\label{eq:DeltaL}
- {\cal L}'=\frac{1}{2\Lambda}\big(\phi^{C\dagger}
\tau_i\phi\big)
\big(\overline{l_R^{0C}}h\,i\tau_2\tau_i l_L^{0}\big) +{\mbox
h.c.}\,,
\end{equation}
with $\tau_i$, $i=1,2,3$ being the Pauli matrices, $h$ being a
$d\times d$ symmetric constant matrix, $\Lambda\gg v$ being the
lepton number violating mass scale (supposedly of order of the
singlet Majorana masses) and $v$  being the Higgs vacuum
expectation value. The  above operator with the effective
isotriplet field
$\Delta_i=(1/\Lambda)\big(\phi^{C\dagger}\tau_i\phi\big)$ reflects
the oblique radiative corrections in the low
energy  Lagrangian produced by the physics beyond
the~SM.\footnote{Were the isotriplet  $\Delta_i$ be
considered as elementaty in the renormalizable framework, it would
change only the  emerging Yukawa interactions
not affecting the mass and mixing matrices.} 
With the Higgs doublet as
\begin{equation}
\phi  = \left( \begin{array}{c} iw^{+} \\
\frac{1}{\sqrt 2}(v+H+iz)
\end{array} \right)\,, 
\end{equation}
it yields (in the 
unitary gauge) the following mass and Yukawa term for neutrinos 
\begin{equation}\label{eq:DeltaL'}
- {\cal L}'= \frac{1}{2}\Big(
1+\frac{H}{v}\Big)^2\,
\overline{\nu_R^{0C}}
\mu\, \nu_L^0 +{\mbox h.c.} \,,
\end{equation}
where $\mu=hv^2/\Lambda $.
Such an effective SM extension will be designated as $(d,s)$. 

Now, let us introduce the complete one-handed neutrino collection
(which can always be chosen, say, as left-handed)
\begin{equation}\label{eq:neutr_set}
n^{\tiny{0}}_L = (\nu_L^{\tiny{0}}, {\nu}^{\tiny{0}C}_L)\,,
\end{equation}
so that $(n^{\tiny{0}}_L)^{\tiny{C}}\equiv
n^{\tiny{0C}}_R = (\nu_R^{\tiny{0C}}, {\nu}^{\tiny{0}}_R)$. 
In these notations, the total neutrino mass matrix ${\cal M}^n_0$ 
defined by the mass Lagrangian
\begin{equation}\label{eq:neutr_mass}
-{\cal L}^n_{mass}=\frac{1}{2}\, \overline{{n}_R^{\tiny{0}C}}{\cal
M}^n_0
{n}^{\tiny{0}}_L+\mbox{h.c.}
\equiv 
-\frac{1}{2}\, n^{\tiny{0}T}_L{} C^{-1} {\cal M}^n_0
n^{\tiny{0}}_L+\mbox{h.c.}
\end{equation}
is clearly symmetric with account for $C^T = -C$. More particularly,
it has the form
\begin{equation}\label{eq:M_nu}
{\cal M}^{n}_0 = 
\left(
\begin{array}{cc} 
\mu &m\\
m^T &M
\end{array}
\right),
\end{equation}
where $m\equiv Y^*v/\sqrt{2}$ 
is an arbitrary $d\times s$ matrix of the Dirac masses,
$m^T$ is its transposed, $\mu$  and $M$ are, respectively,  the
$d\times d$ and  $s\times s$ symmetric Majorana
mass matrices from eqs.~({\ref{eq:lagrangian}) and
({\ref{eq:DeltaL'}).

\paragraph{(ii) Mass  basis} 

Let us now consider  the mass basis  $n_L$ where, by definition,
the neutrino mass matrix is diagonal. 
It is understood in this  
that  the true neutrino mass eigenfields are
$d+s$  four component  fields ${\cal N}(n_L)$ 
bringing the neutrino kinetic Lagrangian to the  diagonal positive
form and simultaneously satisfying  some subsidiary Majorana 
condition to halve the number of degrees of freedom.  
Such  most general 
condition  looks like~\cite{MC}--\cite{bil}
\begin{equation}\label{eq:Maj}
{\cal N}_\l^C\equiv \l\, {\cal N}_\l,
\end{equation}
where  ${\cal N}_\l^C\equiv C{\overline{\cal
N}}_\l^{\,T}$ 
and  $\l=\mbox{diag}\,(\l_1,\dots,\l_{d+s})$ is a diagonal phase
matrix. Here and in what follows we use 
the notations  with subscript
$\l$ to stress that  quantity at hand generally depends on
$\l$.\footnote{Note that the maximum number of the independent 
Majorana specific phases in $\l$ might be $d+s-1$ because
an overall neutrino phase is unobservable.}
Expressing ${\cal N}_\l$   through Weyl spinors as  ${\cal
N}_\l=\l_L{n}_L\oplus \l_R{n}^C_R$ with some diagonal
phase matrices
$\l_L$ and $\l_R$, one finds the Majorana condition to
fulfil if  $ \l_L\l_R=\l^{*}$. Without
loss of generality one can put, e.g.,  $\l_L=I$, $\l_R=\l^{*}$,
so that\footnote{Note that eqs.~(\ref{eq:Maj}), (\ref{eq:N}) do not
put any constraint on  the original Weyl fields $n_L$.}
\begin{equation}\label{eq:N}
{\cal N}_\l={n}_L\oplus \l^{*} {n}^{\tiny C}_R\,. 
\end{equation} 
This choice is advantageous because it results in the simplest form
for the charged current which is left-handed (see later on).

In these terms, we  demand the kinetic part of the
neutrino Lagrangian~be
\begin{equation}\label{eq:L_kin}
{\cal L}_{kin}^n=
\frac{1}{2}\,\overline{\cal N}\!_\l\, i\d   {\cal N}_\l
-\frac{1}{2}\,\overline{\cal N}\!_\l{\cal M}^{n}_{\mbox{\scriptsize
diag}}{\cal N}_\l\,,
\end{equation}
with a  non-negative diagonal mass matrix ${\cal
M}^{n}_{\mbox{\scriptsize diag}}$ idependent of $\l$.
To this end, let us choose  the $(d+s)\times (d+s)$ unitary
transformation
${{\cal U}^n_\l}$ 
\begin{equation}\label{eq:neutr_tr}
n^{\tiny{0}}_L = {\cal U}^n_\l {n}_L\,,
\end{equation}
so that
\begin{equation}\label{eq:neutr_d}
{\cal U}_\l^{nT} {\cal M}^n_0 \,{\cal U}_\l^n =
\l {\cal M}^n_{\mbox{\scriptsize diag}}\,,
\end{equation}
with\footnote{This notations
correspond to  partition ${\cal N}_\l\equiv
(\nu,N)_\l$ and tacitly imply  the see-saw
hierarchy
${m}^\nu \ll {M}{}^N$ for all the elements, with $\nu$ being
(quasi-)doublet neutrinos and $N$ being  (quasi-)singlet ones.
Nevertheless there might be experimental indications of the existence
of at least one light singlet neutrino~\cite{now}.} 
\begin{equation}\label{eq:neutr_d'}
{\cal M}^n_{\mbox{\scriptsize diag}}
= \mbox{diag}\, ({m}^\nu_1, \dots {m}^\nu_d;
{M}{}^N_{1}, \dots {M}{}^N_{s})\,.
\end{equation}
With account for 
$\overline{\cal N\!}_\l=\overline{{\cal N}_\l^C}\l\equiv -{\cal
N}_\l^T
C^{-1}\l $, the neutrino kinetic Lagrangian takes
the required form of eq.~(\ref{eq:L_kin}).
At $s\leq d $ for the $(d,s)_{\mbox{\scriptsize r}}$ extension,
$d-s$
elements ${m}^\nu$  are  zero. This reflects the fact
that in this case the rank of the $(d+s)\times (d+s)$  matrix
given by eq.~(\ref{eq:M_nu}) with $\mu=0$ is  $2s$. At $s> d$ the
rank of the matrix is generally  $d+s$ and hence there is
no massless neutrinos.

Similarly, the charged lepton fields $e_\chi$ ($\chi=L$, $R$) in the
mass basis are defined as
\begin{equation}\label{eq:e_tr}
e^{\tiny{0}}_\chi={\cal U}^e_\chi e_\chi
\end{equation}
with the  unitary $d\times d$ matrices ${\cal U}^e_\chi$, so that the 
bi-diagonalization of relevant mass matrix looks like 
\begin{equation}\label{eq:lep_d}
{{\cal U}^e_L}^\dagger {\cal M}^e_0 \,{\cal U}^e_R = {\cal
M}^e_{\mbox{\scriptsize diag}}
= \mbox{diag}\, ({m}^e_1,  \dots,
{m}^e_{d})\,.
\end{equation}
By means of the global symmetries of the kinetic part of Lagrangian
(\ref{eq:lagrangian})
one can arrange, without  loss
of generality, the  charged lepton weak basis to coincide with the
mass one. This  means
that  ${\cal M}^e_0$ can be chosen diagonal ab initio, so
that ${\cal U}^e_L=  {\cal U}^e_R=I$. 
The associated  neutrinos are
usually referred to as the flavour ones.\footnote{Unfortunatelly,
this is unlike the quark sector where flavour is  synonymous 
with the mass eigenstate.} Traditionally, the corresponding basis is
used when discussing the neutrino oscillation phenomenon. For
simplicity, it is adopted in what follows.
But in fact there is no need for such a
particular choice. Moreover, the mass basis suffices to describe the
neutrino oscillations  without resort to the weak
eigenstates~\cite{giunti}.\footnote{When there is 
an admixture of the heavy Majorana neutrinos, it is only the
coherent part of the light neutrinos what has the meaning
of a flavour state. Note that so modified flavour  states are
non-orthogonal  and process dependent.}

Now, the charged current Lagrangian in the mass basis
reads\footnote{Note that due to the supposed absence of the
vector-like lepton doublets, the  right-handed charged currents do not
emerge.} 
\begin{equation}\label{eq:L_W}
- {\cal L}_W = \frac{g}{\sqrt{2}} W_\alpha^-
                \overline{e_L}\gamma^\alpha { V}_\l {\cal
                N}_{\l L}
                +\mbox{h.c.}\,,
\end{equation}
where the rectangular $ d\times(d+s)$ mixing matrix for the charged
currents is 
\begin{equation}\label{eq:V}
{ V}_\l = {{\cal U}}_L^{e\dagger} P^{en} {\cal U}_\l^n
\end{equation}
with the  charged current matrix  in the weak basis given by
\begin{equation}\label{eq:I_nue}
P^{en} = 
\left(
\begin{array}{cc} 
I_d &\! O_{d\times s}
\end{array}
\right)\,,
\end{equation}
$I_d$ being the $d$-dimensional identity matrix and $O_{d\times s}$
being $d\times s$ zero matrix. 
The lepton mixing matrix ${ V}_\l $ is a counterpart of the quark
CKM matrix. 
It follows from eqs.~(\ref{eq:V}) and (\ref{eq:I_nue}) that 
\begin{equation}\label{eq:genun}
{V}_\l { V}_\l^{\dagger}= I_d\,, 
\end{equation}
though $ V_\l^{\dagger}  V_\l\neq I_{d+s}$. Eq.~(\ref{eq:genun}) can
be regarded as the one-sided unitarity condition at $s\neq 0$.

The neutral current Lagrangian with the SM neutral current operator 
$T_3 - s_W^2 Q$  in the mass basis is as follows:
\begin{equation}\label{eq:L_Z}
- {\cal L}_Z = \frac{g}{c_W} Z_\alpha
\Big{(} - \frac{1}{2}\overline{e_L}\gamma^\alpha e_L
        + s^2_W \overline{e}\gamma^\alpha e 
        + \frac{1}{2} \overline{\cal N}\!_{\l L}\,\gamma^\alpha{
        X}_\l\,
	   {\cal N}_{\l L}
\Big{)}\,,
\end{equation}
where the $(d+s)\times (d+s)$  neutrino mixing matrix for the neutral
currents is 
\begin{equation}\label{eq:X}
{X}_\l  = {\cal U}^{n \dagger}_\l  P^{n}\,  {\cal U}^n_\l 
\end{equation}
with the on-doublet neutrino projector ($P^{n 2}=P^{n}$)
\begin{equation}\label{eq:I_nu}
P^{n} = \mbox{diag}\, (\,\underbrace{1,\dots, 1}_{d};
                         \,\underbrace{0,\dots, 0}_{s}\,)\,.
\end{equation}
Here one puts $c_W\equiv \cos\theta_W$, $s_W\equiv \sin\theta_W$
with $\theta_W$ being the Weinberg angle.
Clearly,    $X_\l $ is a Hermitian projective matrix: 
${X}_\l  ={X}_\l ^\dagger$, ${X}_\l ^2= X_\l \neq I$. Due to
eqs.~(\ref{eq:V}), (\ref{eq:I_nue})  
the relation 
\begin{equation}
{X}_\l ={V}_\l ^\dagger {V}_\l 
\end{equation}
between the neutral and charged current mixing  matrices is
obeyed.
For the  $(d,0)$ extension  one has $P^n=I_d$, so that ${X}_\l \equiv
I_d$ and thus the
$d\times d$ matrix $V_\l $ is unitary. More than this, for the
renormalizable $(d,0)_{\mbox{\scriptsize r}}$ extensions one can
always put ${\cal U}_\l ^n=I$,
so that  $ V_\l = I_d$ also follows.
Hence, the lepton flavour conservation of the minimal SM with the
residual symmetry $U(1)^d$ is readily recovered. 

For the renormalizable extensions
$(d,s)_{\mbox{\scriptsize r}}$, the  Yukawa Lagrangian  looks like 
\begin{eqnarray}\label{eq:L_Y1}
-{\cal L}_{Y}
&=&\phantom{+}\frac{H}{v}\,\overline{e}\,{\cal
M}^e_{\mbox{\scriptsize diag}}e  +    
\frac{z}{v}\,\overline{e}\,{\cal
M}^e_{\mbox{\scriptsize diag}}i\gamma_5\, e   \nonumber\\
&&+\bigg(\frac{1}{2}\frac{H+iz}{v}\,
\overline{\cal N}\!_{\l R } 
\Big(\l^*{X}_\l ^T\l
{\cal M}^n_{\mbox{\scriptsize diag}}+ {\cal M}^n_{\mbox{\scriptsize
diag}}{X}_\l \Big){\cal N}_{\l L}
\nonumber\\
&&+\sqrt{ 2}i\frac{w^- }{v\,\,}\Big(\overline{e_L}\,V_\l
{\cal M}^n_{\mbox{\scriptsize diag}}{\cal N}_{\l R }-\overline{e_R}
{\cal M}^e_{\mbox{\scriptsize diag}}V_\l{\cal N}_{\l L}\Big)
+\mbox{h.c.}\bigg).
\end{eqnarray}
Here use is made of the constraint
\begin{equation}\label{eq:constraint'}
{X}_\l ^T \l {\cal M}^{n}_{\mbox{\scriptsize diag}}{}{X}_\l =0\,,
\end{equation}
which follows from a more particular one 
\begin{equation}\label{eq:constraint}
P^{n}\, {\cal U}_\l ^n{}^* \l{\cal
M}^{n}_{\mbox{\scriptsize diag}}\,{\cal
U}_\l ^n{}^\dagger P^{n}=0
\end{equation}
and reflects the absence of the $d\times d$
symmetric  left-handed Majorana
mass term $\mu$ in eq.~(\ref{eq:M_nu}).

For the  general extensions $(d,s)$, the constraint
eq.~(\ref{eq:constraint'}) should be dropped off. This results in
addition of a number of  interaction  terms to
Yukawa Lagrangian. E.g., according to eq.~(\ref{eq:DeltaL'})  one
should add in the unitary gauge  the term
\begin{equation}\label{eq:L_Y2}
-{\cal
L}'_Y=\frac{1}{2}\Big(\frac{H}{v}\Big)^2\,
\overline{\cal
N}\!_{\l R }\,
\l^*{X}_\l ^T \l {\cal M}^n_{\mbox{\scriptsize diag}}{X}_\l 
{\cal N}_{\l L}+
\mbox{h.c.}\,,
\end{equation}
the linear in $H$ term being
cancelled by a similar one   present now in~${\cal L}_Y$. 

\section{Rephasing invariance}

Consider the group of transformations consisting of the
Majorana field rephasing   ${\cal N}_\l\to \Phi^{1/2}{\cal N}_\l$
followed by transformations 
\begin{eqnarray}\label{eq:RI}
\l&\to&\l \Phi^*,\nonumber\\
{V}_\l &\to&{V}_\l \Phi^{*1/2}
\end{eqnarray}
with a diagonal phase matrix
$\Phi=\mbox{diag}\,(\Phi_1,\dots,\Phi_{d+s})$. As a result, one also
gets ${X}_\l \to \Phi^{1/2} {X}_\l \Phi^{*1/2}$. 
All the Lagrangians are clearly rephasing invariant.
It follows from eq.~(\ref{eq:RI}) that independent  rephasing
invariant
quantities  containing $\l$ may be chosen as $V_\l \l^{*1/2}$ 
($\l^{1/2}X_\l \l^{*1/2}$)
and  $\l^{1/2}{\cal N}_\l$. Thus, the rephasing  allows one to
extract a number of  neutrino phases from ${V}_\l$  and to
reabsorb them in $\l$ (or v.v.). Observables  depend only on
the sum of the complementary 
phases of ${V}_\l$   and 
$\l^{*1/2}$ as well as of ${\cal N}_\l$  and  $\l^{1/2}$, but not
separately on each of them (in addition to phases in the rephasing
invariant combinations of
the matrix $V_\l$ itself).
Clearly, it is not a particular choice of the Majorana condition but
the invariance with respect to this choice  which is 
physically meaningful.\footnote{Stress that due to
rephasing invariance, fixing a choice for $\l$ has 
nothing to do with the real physical
properties of the Majorana neutrinos, in particular with those
concerning  $C$ conjugation. 
The last properties are described additionally 
by the fact that if  the neutrino mass eigenstates  do possess
definite $C$ parity $\eta_C=\mbox{diag}\,(\pm1)$,  then the $C$
conjugation  for the Majorana eigenfields shoud be
consistently redefined~\cite{kaiser} as  
${\cal N}_\l\stackrel{C}{\to}{\cal N}_\l^{C_\l}\equiv\eta_C\l^*{\cal
N}_\l^C$, where traditionally ${\cal N}_\l^C\equiv C{\overline{\cal
N}}_\l^{\,T}$. 
It follows that the modified (anti-)self-charge conjugacy condition
${\cal N}_\l^{C_\l}=\eta_C {\cal N}_\l$ is indeed satisfied
independent of $\l$. 
Nevertheless, attempts are sometimes made in the
literature to ascribe physics content to
the Majorana condition being chosen superficially 
in the  self- or anti-self-charge conjugate form. 
The emerging results  are thus misleading.} 

The rephasing invariance permits one to choose $\l$
most appropriate to the problem at hand. The reason is that
only  the Higgs vertices and  the neutrino wave functions
(and thus the $<\!\!{\cal N}_\l\overline{{\cal N}_\l^C}\!\!>$
propagators)  depend explicitly on~$\l$, whereas the
gauge vertices  and  the 
$<\!\!{\cal N}_\l\overline{{\cal N}_\l}\!\!>$ propagators
do not depend on it. As a result,
if the  matrix element for  a particular process does not
contain $\l$ explicitly  one can extract by means of the rephasing as
many Majorana specific phases from $V_\l$ as possible.  The  rephasing
invariance then
insures that under other $\l$ these phases, though being superficially
present in $V_\l$, would
not enter nevertheless  the final results.

To illustrate, the amplitude for the (chirality conserving)
$\cal NN$ oscillations 
\begin{equation}\label{eq:A0}
{\cal A}_0(t)=V_\l e^{-iEt} V_\l^\dagger
\end{equation}
clearly does not depend on the Majorana specific phases capable of
being stored in $\l$, whereas amplitude
for the  (chirality flipping)  ${\cal NN}^C$  oscillations 
\begin{equation}\label{eq:A1}
{\cal A}_1(t)= V_\l e^{-iEt} \l^* {\cal M}^n_{\mbox{\scriptsize
diag}}E^{-1} V_\l^T 
\end{equation}
does depend on the phases. In the above,  $E$ is the diagonal energy
matrix for the
light neutrinos. The same is true for the  neutrino mass
elements $\nu\nu^C$ in the weak basis
\begin{equation}\label{eq:2B}
{\cal M}^{n\,*}_{\nu_e\nu_{e'}}=\big(V_\l \l^*{\cal
M}^n_{\mbox{\scriptsize diag}}V_\l^T\big)_{\nu_e\nu_{e'}}\,,
\end{equation}
which determine the rates of the neutrinoless double $\beta$-decay
(at $e'=e$) or $e\bar\mu$ conversion (at $e'=\mu$).\footnote{Note that
according to eqs.~(\ref{eq:L_Y1}) and  (\ref{eq:L_Y2}) 
the (chirality flipping) Yukawa interactions  might also serve as a
probe of the Majorana specific phases.}

\paragraph{(i) Canonical Majorana condition}

Sometimes it might be tempting to go to a basis where the Majorana
neutrino
wave functions have a canonical form. Namely,  the rephasing by
$\Phi=\l$ yields   $\l\to I$, with $I$ 
being unity matrix, and transformed  fields ${\cal N}_I$ satisfy the
canonical
Majorana condition ${\cal N}_I^C={\cal N}_I$.
Under this  condition, all the  physical mixing
parameters reside only in mixing matrices.    
With account for the $X_\l$  Hermiticity property
${\cal R}e\,{X}_\l^T={\cal R}e\,{X }_\l$ and
${\cal I}m\,{X}_\l^T=-{\cal I}m\,{X}_\l$, the neutrino neutral current 
parts of Lagrangians (\ref{eq:L_Z}) and
(\ref{eq:L_Y1})
can now be re-expressed  in a simpler form 
\begin{equation}\label{eq:L_Z1}
- {\cal L}_Z^n = \frac{g}{4 c_W} Z_\alpha\,
\overline{\cal N}{\!_I}\gamma^\alpha \Big(i\, {\cal I}m\,{X}_I\,
 -\gamma_5\, { \cal R}e\, {X}_I\, \Big)\, {\cal N}{\!_I}
\end{equation}
and\footnote{Under $\l= I$, the Yukawa term for the renormalizable
extensions $(n,n)_{\mbox{\scriptsize r}}$ was found in
ref.~\cite{pilaft}.} 
\begin{eqnarray}\label{eq:L_Y'}
- {\cal L}_{Y}^n &=& 
\phantom{+}\frac{1}{2} \frac{H}{v}\overline{\cal N}{\!_I}\Big(
{\cal R}e\,{X}_I+i\gamma_5\,{\cal I}m\,{X}_I  \Big){\cal
M}_{\mbox{\scriptsize diag}}^n\, 
{\cal N}{\!_I}\nonumber\\
&&+\frac{1}{2} \frac{z}{v}\,\overline{\cal N}{\!_I}\,
\Big( {\cal I}m\,  {X}_I
-i\gamma_5\,{\cal R}e\,{X}_I  \Big){\cal M}_{\mbox{\scriptsize
diag}}^n\, {\cal N}{\!_I}
+\mbox{h.c.}
\end{eqnarray}

\paragraph{(ii) $CP$ invariance}

It is well known that for a field theory
not to explicitly violate  $CP$  there should be allowed a weak
basis where all the parameters in the Lagrangian are real. Under this
condition, the neutrino mass matrix
${\cal M}^n$, being symmetric,  can always be brought to
the (real) diagonal form (generally, not positive definite)  
by means of an  orthogonal transformation ${\cal U}^n= {\cal O}^n$
with the effect 
\begin{equation}
{\cal O}^{nT}\!{\cal M}^n{\cal O}^n=\eta_M {\cal
M}^n_{\mbox{\scriptsize diag}}\,.
\end{equation}
Here $\eta_{M}\equiv\mbox{diag}\,(\pm1)$ is the mass signature
matrix which is completely determined by the original  ${\cal M}^n$.
Clearly, the mixing matrix $V_{\eta_M}\equiv {\cal
R}=P^{en}{\cal O}^n$ is real.\footnote{Here and in what follows,
the basis where  ${\cal U}^e_L=  {\cal U}^e_R=I$ is generally  
chosen
for simplicity.}
In the rephasing invariant form  one gets ${V}_\l={\cal
R}\,(\l\eta_{M})^{1/2}$, and hence the condition for $CP$ invariance
looks like 
\begin{equation}\label{eq:eta}
{V}_\l= {V}_\l^* \l\eta_{M}
\end{equation}
as well as
\begin{equation}\label{eq:X'}
{X}_\l=  \eta_{M}\l^*{X}_\l^* \l\eta_{M}\,.
\end{equation}
Stress that CP conservation does not mean ${V}_\l$ and ${X}_\l$ to be
real in general. 

In the mass basis, one can define the  $CP$
conjugation (in the unitary gauge) as 
\begin{eqnarray}\label{eq:CP}
e(x)&\to &\gamma_0e^C(x^P)\,,\nonumber\\
{\cal N}_\l(x) &\to & \eta_{CP}\l^*\gamma_0 {\cal 
N}_\l^C(x^P)\,,\nonumber\\ 
W^{\pm}(x)&\to &- W^{\mp\, P}(x^P)\,,\nonumber\\
Z(x)&\to & -Z^P(x^P)\,,\nonumber\\
H(x)&\to & H(x^P)
\end{eqnarray}
(with $x^P\equiv
(x_0, -\stackrel{\to}{x})$, etc). The definition  is clearly rephasing
invariant. 
Here  $i\eta_{CP}$, with $\eta_{CP}\equiv\mbox{diag}\,(\pm1)$, 
is the matrix of the (relative) $CP$ parities for the neutrino
mass eigenstates~\cite{kaiser,bil}.  In the above, $\eta_{CP}$ is
not arbitrary but is to be properly defined  for consistency. 
Namely, under eq.~(\ref{eq:CP}) the whole Lagrangian
can be shown to transform into itself where substitutions 
\begin{eqnarray}\label{eq:CP'}
\l&\to&\l\,,\nonumber\\
V_\l&\to&V_\l^* \l\eta_{CP}
\end{eqnarray}
are made. Imposing the requirement of $CP$ invariance one arrives with
account for eq.~(\ref{eq:eta}) at the identity
\begin{equation}\label{eq:CPM}
\eta_{CP}\equiv\eta_{M}\,.
\end{equation} 
This identity insures the consistency  of the description of $CP$
invariance  directly in terms of the rephasing invariant quantities, 
which being  built of $V_\l$ and $\l$ depend on $\eta_{M}$, 
with the description in terms of the explicit $CP$ transformations 
eq.~(\ref{eq:CP}) being dependent on  $\eta_{CP}$.  

In particular, in the case of CP conservation one gets for
the  amplitudes of eqs.~(\ref{eq:A0})--(\ref{eq:2B})
\begin{eqnarray}
{\cal A}_0(t)&=&V_\l e^{-iEt} \eta_{CP}\,\l^* V_\l^T\,,\nonumber\\
{\cal A}_1(t)&=&V_\l e^{-iEt} \eta_{CP}{\cal M}^n_{\mbox{\scriptsize
diag}}E^{-1}
V_\l^\dagger\,,\nonumber\\
{\cal M}^{n\,*}_{\nu_e\nu_{e'}}&=&
\big(V_\l\,\eta_{CP} {\cal M}^n_{\mbox{\scriptsize diag}}
V_\l^\dagger\big)_{\nu_e\nu_{e'}}\,.
\end{eqnarray}
At $\nu_e=\nu_{e'}$, the last line explicitly demonstrates  the
possibility for the 
(partial) compensation of  various 
contributions to the lepton number violating $e\bar e$
transition under $CP$ conservation.
At  $\l=\eta_{CP}$, the matrix $V_\l$ (as well as  $X_\l$)
becomes pure real, $V_{\eta_{CP}}\equiv \cal R$, so that
\begin{eqnarray}
{\cal A}_0(t)&=&{\cal R}e^{-iEt} {\cal R}^T\,,\nonumber\\
{\cal A}_1(t)&=&{\cal
R}e^{-iEt} \eta_{CP}{\cal M}^n_{\mbox{\scriptsize diag}}E^{-1}{\cal
R}^T\,,\nonumber\\
{\cal M}^n_{\nu_e\nu_{e'}}&=&
\big({\cal R}\eta_{CP} {\cal M}^n_{\mbox{\scriptsize diag}}{\cal
R}^T\big)_{\nu_e\nu_{e'}}\,.
\end{eqnarray}

The   basis ${\cal N}_{\eta_{CP}}$ may be called as the
$CP$-associated one.  In a sense, it might present the most natural
choice for the $CP$  conserving theory,  all other bases being
equivalent though  probably less convenient.
Thus, under canonical
Majorana condition $ \l=I$ the elements of $V_I$ (and $X_I$) in the
$CP$ conserving theory should be according to eq.~(\ref{eq:eta}) 
either pure real or imaginary~\cite{kaiser,bil,CP}, and this has
nothing
to do with the maximal $CP$ violation as it might superficially seem.

\section{Doublet-singlet parametrization}

\paragraph{(i) General case}

A  mathematical parametrization of the neutrino mixing matrix
${\cal U}^n$ is given in ref.~\cite{nm}. 
An alternative physical prescription, heavily relying on  the
doublet-singlet neutrino content  and thus being useful for practical
purposes, is  proposed in the present paper. For simplicity,  the
subscript $\l$ will be  omitted in what follows. First of all note
that by means of the global symmetries  one can always achieve,
without loss of generality, that ${\cal U}^e_L={\cal
U}^e_R=I$.\footnote{For
this reason, lepton mixing is synonimous with the neutrino one.} Now, 
before applying  any restrictions on the neutrino mass matrix ${\cal
M}^n$ the $(d+s)\times (d+s)$ unitary mixing  matrix ${\cal U}^n$
is  arbitrary and  can be  decomposed in a unique way (at least in a
neighborhood of unity) as
\begin{equation}\label{eq:U}
{\cal U}^n={\cal U}^n_d\,  {\cal U}^n_s \,{\cal U}^n_m\,. 
\end{equation}

Here
${\cal U}^n_d$ is a unitary $d\times d$ matrix in the doublet
neutrino subspace corresponding to  indices $f=1, \dots, d$.  This
matrix is spanned on  $d^2$ generators and depends on $d(d-1)/2$
mixing angles and $d(d+1)/2$ phases. More particularly, one can put
\begin{equation}\label{eq:U_d1}
{\cal U}^n_d =
\left (
\begin{array}{cc}
U^\nu_d&0\\
0&I_s
\end{array}
\right)
\end{equation}
with a $d\times d$ unitary matrix $U^\nu_d$.
There is still a freedom of $d$
charged lepton phase
redefinition which is left  after  the  mass matrix in
eq.~(\ref{eq:lep_d}) is  diagonalized. According to 
eqs.~(\ref{eq:V})  and (\ref{eq:U}) this freedom can be used to
eliminate $d$ phases out of ${\cal U}^n_d$.  
It clearly leaves  only  $d(d-1)/2$
independent phases  in this matrix (and  equal number of 
mixing angles).

Now, one can write down the following explicit parametrization
for $U_d$ ($d>1$) in terms of the modified Pontryagin's coordinates of
the second kind~\cite{nm}
\begin{equation}\label{eq:U_par}
U^\nu_d=u_{\mbox{\scriptsize
diag}}(\alpha)\prod_{\stackrel{f,g=1,\dots,d}{f<g}}
\otimes\,
u_{fg}(\theta,
\delta)\,.
\end{equation}
The product above
should be understood in some particular (but a priori unspecified)
order.
Here $u_{\mbox{\scriptsize diag}}$ is a  diagonal $d\times d$ phase
matrix 
$u_{\mbox{\scriptsize diag}}(\alpha)=\mbox{
diag}\,(e^{i\alpha_1},\dots,e^{i\alpha_d})$
which differs equivalent parametrizations and is at our disposal. 
(At $d=1$, one has $U^\nu_1=e^{i\alpha_1}$.)
A basic matrix $u_{fg}$ (``complex rotation''), one of a set of
$d(d-1)/2$ unitary $SU(2)$ submatrices, acts in the $fg$
plane, $f\ne g$, and 
depends only on one mixing angle $\theta_{fg}$ and one
phase~$\delta_{fg}$
\begin{equation}\label{eq:u_fg}
u_{fg}=
\exp \left(
\begin{array}{cc}
0&\theta_{fg} e^{i\delta_{fg}}\\
-\theta_{fg} e^{-i\delta_{fg}}&0
\end{array}\right)
=
\left(
\begin{tabular}{cc}
$\cos\theta_{fg}$&$\sin\theta_{fg}e^{i\delta_{fg}}$\\
$-\sin\theta_{fg}e^{-i\delta_{fg}}$&$\cos\theta_{fg}$
\end{tabular}
\right)\,.
\end{equation}
By means of the identity
\begin{equation}\label{eq:identity}
u_{\mbox{\scriptsize diag}} (\alpha)\,
u_{fg}(\theta_{fg},\delta_{fg})\,
u^{\dagger}_{\mbox{\scriptsize diag}}(\alpha)
=u_{fg} (\theta_{fg},\alpha_f+\delta_{fg}-\alpha_g)
\end{equation} one can eliminate 
$d-1$ $\delta$'s out of $U^\nu_d$ and
to transform these phases into the same number of the Majorana
specific  ones, the $d$-th of the last phases being
unphysical.\footnote{Strictly
speaking, this is  true only for the $(d,0)$ case. For the $(d,s)$
extension the Majorana specific phases could be exposed only after
taking into account the  matrix ${\cal U}^n_m$.} It clearly leaves
$(d-1)(d-2)/2$ CKM-like phases and $d-1$ Majorana specific ones. 
Thus, under  proper phase redefinitions the matrix $U^\nu_d$ may be
chosen in experimentally viable cases $d=2$ and 3, respectively, as
\begin{equation}\label{U_2}
U^\nu_2=
\left(
\begin{array}{cc}
c&s\\
-s&c
\end{array}
\right)\left(
\begin{array}{cc}
e^{i\gamma}&0\\
0&1
\end{array}\right)
\end{equation}
and
\begin{eqnarray}\label{U_3}
U^\nu_3&=&
\left(
\begin{array}{ccc}
c_3&s_3&0\\
-s_3&c_3&0\\
0&0&1
\end{array}
\right)\left(
\begin{array}{ccc}
c_2&0&s_2\\
0&1&0\\
-s_2&0&c_2
\end{array}\right)\\\nonumber
&\times&\left(
\begin{array}{ccc}
1&0&0\\
0&c_1&s_1e^{i\delta}\\
0&-s_1e^{-i\delta}&c_1
\end{array}
\right)\left(
\begin{array}{ccc}
e^{i\gamma_1}&0&0\\
0&e^{i\gamma_2}&0\\
0&0&1
\end{array}
\right), 
\end{eqnarray}
where $c\equiv \cos\theta$ and $s\equiv \sin\theta$. Clearly, one
can shift the ordinary phase  $\delta$ to any of the $s_i$, 
$i=1, 2, 3$.

Further, ${\cal U}^n_s$ is  the 
counterpart  of ${\cal U}^n_d$ in the singlet neutrino
subspace with  indices $f=d+1, \dots, d+s$, being spanned on  the
$s^2$ generators and dependent on $s(s-1)/2$
mixing angles and $s(s+1)/2$ phases. One has
\begin{equation}\label{eq:U_s1}
{\cal U}^n_s =
\left (
\begin{array}{cc}
I_d&0\\
0&U^N_s
\end{array}
\right)
\end{equation}
with a $s\times s$ unitary matrix $U^N_s$.
Clearly, ${\cal
U}^n_d$ and  ${\cal U}^n_s$  commutes with each other. 
According to eq.~(\ref{eq:V}) the matrix ${\cal U}^n_s$ is irrelevant
for observables. Hence, by means of the global
symmetries  one can always achieve, without  loss of generality,
that  $U^N_s=\mbox{diag}\,(e^{i\alpha_{d+1}},\dots,
e^{i\alpha_{d+s}}$),
with $\alpha$'s being at our disposal. This choice is advantageous to
subsequently expose the Majorana specific phases in ${\cal U}^n$.  

Finally,  ${\cal U}^n_m$
is a unitary $(d+s)\times (d+s)$ matrix spanned on $2sd$ 
generators which mix the two subspaces.\footnote{This follows from
inversion of eq.~(\ref{eq:neutr_tr}) for transformation between the
weak and mass neutrino bases.} This matrix depends generally
on $sd$ mixing angles and the same number of  phases. 
It follows from  eqs.~(\ref{eq:X}) and  (\ref{eq:U}) that the neutral
current mixing matrix takes the form
\begin{equation}\label{eq:X_m}
{X} = {\cal U}_m^{n \dagger} P^{n} \, {\cal U}_m^n\,.
\end{equation}
In other words, it depends entirely on the  parameters of ${\cal
U}_m^n$, the rest  of  parameters  present in ${\cal
U}^n_d$ manifesting themselves only through
charged currents (and thus through neutrino oscillations).
To achieve this goal, the chosen order of matrices ${\cal U}_d^n$ and 
${\cal U}_s^n$ relative to  ${\cal U}_m^n$ in eq.~(\ref{eq:U})  is
crucial.
The  factorization property of the charged and neutral currents
makes the parametrization eq.~(\ref{eq:U}) very convenient in
practice.  Altogether,
the total neutrino mixing matrix 
${\cal U}^n$ for the general $(d,s)$ extension contains 
$d(d-1)/2 +sd$ physical mixing angles and the
same number of phases in  agreement with refs.~\cite{nu1}, \cite{nm}. 
Similarly to eqs.~(\ref{eq:U_par}), (\ref{eq:u_fg}) one can propose
the following
explicit representation for ${\cal U}^n_m$
\begin{equation}\label{eq:U_par'}
{\cal U}^n_m=\prod_{\stackrel{f=1,\dots,d}{g=1,\dots,s}}
\otimes\,
u_{f,d+g}(\omega_{f,d+g})\,
\end{equation}
with a fixed but a priori unspecified order of submatrices, and
$\omega_{f,d+g}$ being $d s$ arbitrary complex numbers. When
restricted to $2\times 2$ complex plane the matrices $u_{f,d+g}$ are
quite similar to those given by eq.~(\ref{eq:u_fg}). 
By means of the identity~(\ref{eq:identity}) with  diagonal phases
from  $U^\nu_s$, one can eliminate  $s$
phases out of $ds$ ones in ${\cal U}^n_m$, and to get in the end
$d+s-1$  Majorana specific phases in~${\cal U}^n$.

As for the renormalizable $(d,s)_{\mbox{\scriptsize r}}$ extensions,
the  $d\times d$ symmetric matrix constraint
eq.~(\ref{eq:constraint})
reduces   $d(d+1)/2$ phases and the same number of moduli, $d$ of
the latter ones  corresponding to  masses and  $d(d-1)/2$ to mixing
angles. As a result,  ${\cal U}^n$
contains $sd$ independent physical mixing angles and $d(s-1)$
phases, precisely as it  should according to general counting of
ref.~\cite{nu1}. Superficially, 
the above constraint restricts
only parameters in ${\cal U}^n_m$ and does not touch those in
$U^\nu_d$.
But it can be shown that at $d\geq s>0$, due to the presence of $d-s$
massless neutrinos,  it is additionally possible to eliminate from
$U^\nu_d$ the parameters corresponding to  $U(d-s)$.   
It leaves in $U^\nu_d$ $ds-s(s+1)/2$ independent $\theta$'s and
$d(s-1)-s(s-1)/2$ $\delta$'s. 
Note that the constraint does not invalidate the
charged-neutral current factorization property.

This gives the complete solution to the problem.
There are two important cases with  neutral currents remaining
diagonal.

\paragraph{(ii) Only Dirac masses}

For the particular case of the $(d,s)_{\mbox{\scriptsize r}}$
extension with only Dirac masses, further reduction
of parameters is possible. Diagonalization of the neutrino mass matrix
by ${\cal U}^n_d\, {\cal U}^n_s$ yields for the
$d\times s$ Dirac mass term 
\begin{equation}
m_{\mbox{\scriptsize diag}}=U^{\nu T}_d m\, U^N_s\,,
\end{equation}
with the non-negative elements on the quasi-diagonal, the rest 
being zero. Remind that $U^N_s$ is unobservable.
At $0<s\leq d$ one has $s$  nonzero entries in $m_{\mbox{\scriptsize
diag}}$. Hence,
there is the $U(d-s)\times U(1)^{s-1}$
left-out symmetry in the doublet neutrino subspace which reduces the
number of parameters in $U^\nu_d$
to $sd -s(s+1)/2$ mixing angles and $sd -s(s+1)/2 -d+1$ phases.
At $0<d< s$ there are  $d$ nonzero entries, the left-out
symmetry in the doublet neutrino subspace is only $U(1)^{d-1}$ and
one recovers
the  CKM-like scheme for $d$ Dirac neutrinos  with 
$d(d-1)/2$ mixing angles and $(d-1)(d-2)/2$ phases.
This explicit counting is  in complete accordance with the general one
in ref.~\cite{nu1}.  

Finally,  there still 
remains the maximal (equal to $\pi/4$) mixing  ${\cal U}^n_m$
between the pairs  of the mass degenerate eigenfields.  Under  the
proper choice for $\l$, the ensuing
orthogonal transformation ${\cal O}^n$ brings the neutrino 
mass matrix to the real diagonal  form  
${\cal M}^n_{\mbox{\scriptsize
diag}}=\big(m_1(1,-1),\dots,m_p(1,-1),0,\dots,0\big)$,
with $p=\mbox{min}\,(d,s)$.  It
corresponds 
to $p$ pairs of the mass degenerate Majorana neutrinos with opposite
$CP$ parities plus $|s-d|$ massless  neutrinos. The  emerging
mixing matrix $X$ in eq.~(\ref{eq:X_m}) is superficially
non-diagonal. Nevertheless the
neutral currents may be put to  explicitly flavour conserving 
form independent of $\l$ via the reversed transition  to the Dirac
basis. As for massless neutrinos, there
is no difference  whether they are considered as Weyl or Majorana
ones.  The  neutral current Lagrangian ${\cal L}_Z^n$  for the doublet
massless neutrinos is flavour conserving, singlet massless neutrinos
being  sterile.

\paragraph{(iii) Only Majorana masses}

In the case of $(d,s)$ extension with only Majorana masses, one has
${\cal U}^n_m\equiv I$ and  hence ${X}\equiv P^{n}$. The neutrino part
of  interactions now becomes 
\begin{equation}
 {\cal L}_Z^n = \frac{g}{4 c_W} Z_\alpha\,
\overline{{\cal N}}\gamma^\alpha\gamma_5 P^{n} {\cal N}
\end{equation}
and (in the unitary gauge)
\begin{equation}
-{\cal
L}_Y^n=\bigg(\frac{H}{v}
+\frac{1}{2} \Big(\frac{H}{v}\Big)^2\bigg)\,
\overline{\cal N}{\cal M}^n_{\mbox{\scriptsize diag}}P^{n}{\cal N}\,,
\end{equation}
both Lagrangians being explicitly independent of $\l$. Due to presence
of the on-doublet neutrino projector $P^n$  the singlet
neutrinos are insured to be sterile.

\paragraph{({iv}) Small doublet-singlet mixing}

It is instructive to discuss the mixing matrices under condition of a
small doublet-singlet mixing, the case  of  importance for
phenomeno\-logy. In particular this is so in the framework of the
see-saw approximation (see further on). 
Making use of the equivalent representation for eq.~(\ref{eq:U_par'})
as 
\begin{equation}\label{eq:U_m1}
{\cal U}^n_m =
\exp\left (
\begin{array}{cc}
0&\kappa\\
-\kappa^{\dagger}&0
\end{array}
\right)\,,
\end{equation}
where $\kappa$ is an arbitrary complex $d\times s$ matrix,
one gets for small $\omega$
\begin{equation}\label{eq:U_m2}
{\cal U}^n_m =
\left (
\begin{array}{cc}
\big(1-\frac{1}{2}\kappa\kappa^\dagger\big)&
\kappa\big(1-\frac{1}{6}
\kappa^\dagger\kappa\big)\\[0.5ex]
-{\kappa^\dagger}\big(1-\frac{1}{6}
\kappa\kappa^{\dagger} \big)
&1-\frac{1}{2}\kappa^\dagger\kappa
\end{array}
\right) +{\cal O}(\kappa^4)
\end{equation}
and 
\begin{equation}\label{eq:U_2}
{\cal U}^n =
\left (
\begin{array}{cc}
U^\nu_d\big (1-\frac{1}{2}\kappa\kappa^\dagger\big)&
U^\nu_d\kappa\big(1-\frac{1}{6}
\kappa^\dagger\kappa\big)\\[0.5ex]
-{\kappa^\dagger}\big(1-\frac{1}{6}
\kappa\kappa^{\dagger} \big)
&1-\frac{1}{2}\kappa^\dagger\kappa
\end{array}
\right) +{\cal O}(\kappa^4)\,.
\end{equation}
Hence one has 
\begin{equation}\label{eq:K}
{V} =
\left(
\begin{array}{cc}
U^\nu_d\,\big(1-\frac{1}{2}\kappa\kappa^\dagger\big)&U^\nu_d\,\kappa
\end{array}
\right) 
+{\cal O}(\kappa^3)\,,
\end{equation}
as well as
\begin{equation}\label{eq:P}
{X} =
\left (
\begin{array}{cc}
1-\kappa\kappa^\dagger&\kappa\\
\kappa^\dagger&\kappa^\dagger\kappa
\end{array}
\right) 
+{\cal O}(\kappa^3)\,.
\end{equation}
These expressions can  readily be generalized with any
finite accuracy in~$\kappa$.\footnote{Clearly, the
above results are not applicable in the case of the pseudo-Dirac
neutrinos where $\kappa$'s are generally  not small. Here 
eq.~(\ref{eq:U_m1}) could be properly
modified by decomposing mixing matrix ${\cal U}^n_m$ into the product
of two parts, 
${\cal U}^n_m\equiv {\cal U}^{n}_{m2}{\cal U}^{n}_{m1}$. The part
${\cal U}^{n}_{m1}$ should  produce transition to the pseudo-Dirac
basis by
a set of the (mutually commuting) pairwise transformations at the
(nearly) $\pi/4$ angles.
The   part ${\cal U}^{n}_{m2}$ due to the rest of $\omega$'s could
result in the remaining flavour violating corrections.}

Finally, the  constraint for the $(d,s)_{\mbox{\scriptsize r}}$
extension given by eq.~(\ref{eq:constraint'})  yields
\begin{equation}\label{eq:constraint''}
\l^{\nu*}m^{\nu}_{\mbox{\scriptsize diag}}=-\kappa
\l^{N*}{M}^{N}_{\mbox{\scriptsize diag}}\kappa^T +{\cal
O}(\kappa^4)\,.
\end{equation}
This determines  $m^{\nu}_{\mbox{\scriptsize diag}}$ and a part of the
$\omega$'s in
terms of $M^N$ and the rest of the $\omega$'s. E.g., in the
simplest case  $s=1$ the solution to the equation can be shown to be
given  by
the $d$-dimensional vector $\omega$ with one nonzero  component
$\omega_d=(-\l_d^\nu/\l^N)^{1/2}|\omega|$, so that $m_d=|\omega|^2M$.
Reversing, one gets generically
$\kappa={\cal O}(|\,{ m}^\nu_{\mbox{\scriptsize
diag}}/{M}{}^N_{\mbox{\scriptsize diag}}|^{1/2})$.
The general solution to eq.~(\ref{eq:constraint''})  is
given by an $s\times s$ nonzero matrix with the proper constraints
followed from the equation. As a result, the parameters in
${\cal U}^n$ are   shared between the independent
ones in ${\cal U}^n_d$ and ${\cal U}^n_m$ as is shown in Table~1. 
\begin{table}[htbp]
\vspace{-2ex}
\paragraph{Table~1} 
Independent mixing parameters for the renormalizable
$(d,s)_{\mbox{\scriptsize r}}$ extensions.
\vspace{1ex}
\begin{center}
\begin{tabular}{|c|c|c|c|c|}
\hline 
&Param's&${\cal U}^n$&${\cal U}^n_d$&${\cal U}^n_m$\\
\hline
&Angles&$ds$&$d(d-1)/2$&$s(s+1)/2$\\ 
$d\geq s>0$&&&$-(d-s)(d-s-1)/2$&\\
\cline{2-5}
&Phases&$d(s-1)$&$d(d-1)/2$&$s(s-1)/2$\\ 
&&&$-(d-s)(d-s+1)/2$&\\
\hline
$s>d>0$&Angles&$ds$&$d(d-1)/2$&$sd-d(d-1)/2$\\ 
\cline{2-5}
&Phases&$d(s-1)$&$d(d-1)/2$&$sd-d(d+1)/2$\\ 
\hline
$d=s=n$&Angles&$n^2$&$n(n-1)/2$&$n(n+1)/2$\\ 
\cline{2-5}
&Phases&$n(n-1)$&$n(n-1)/2$&$n(n-1)/2$\\ 
\hline
\end{tabular}
\end{center}
\end{table} 
The relations above  have their close counterparts in
the framework of the see-saw approximation (see below).  

The part ${\cal U}^n|_{d\times d}$ of  the total mixing matrix
${\cal U}^n$
which spans  the $d\times d$ subspace of the  doublet
neutrinos  reads
\begin{equation}\label{eq:U_light}
{\cal U}^n|_{d\times d}
=U^\nu_d\Big(1-\frac{1}{2}\kappa\kappa^\dagger\Big)+{\cal
O}(\kappa^4)\,.
\end{equation}
It includes the $d\times d$ Hermitian  combination
$\kappa\kappa^\dagger$ of the $d\times s$ matrix  $\kappa$. 
This brings in the additional mixing angles and phases. But even in
neglect of these terms,
when ${\cal U}^n|_{d\times d}=U^\nu_d$ is unitary, the number
of physical phases  in it  being  relevant for the Majorana 
neutrinos, $d(d-1)/2$, would exceed 
that $(d-1)(d-2)/2$  given by the  CKM-like
unitary matrix for the Dirac neutrinos. In
essence, this difference is due to having the freedom of
fixing in $U^\nu_d$, out of the initial $d(d+1)/2$ phases, only
$d$ phases in the Majorana case,
instead of $2d-1$ ones in the Dirac case.

\paragraph{(v) See-saw approximation}

In order to evaluate the mixing magnitudes and study the decoupling
limit, it is useful to compare the general results for small mixing
with those
obtained in the framework of the see-saw mechanism by  the explicit
diagonalization of the neutrino mass matrix. 
By  the unitary global transformation $U(s)$ of the  singlet neutrinos
the mass matrix $M$ in
eq.~(\ref{eq:M_nu}) can be put to the  diagonal form 
\begin {equation}
M=\l^N M^N_{\mbox{\scriptsize diag}}\,.
\end{equation}
Besides, $d$ phases of the Dirac mass matrix $m$ can be eliminated
due to the freedom of the charged lepton phase redefinitions. This
freedom is  still left after the simultaneous diagonalization of the
charged lepton
mass matrix by the bi-unitary $d\times d$
transformation.  So,
the total neutrino mass matrix ${\cal M}^n$ clearly contains
$s(d+1)$ independent moduli, $s$ of them corresponding to 
physical masses and $sd$ ones to mixing angles, as well as
$d(s-1)$
phases. This explicit counting for the $(d,s)_{\mbox{\scriptsize
r}}$ extension is in accordance with the general one presented in
ref.~\cite{nu1}. 

The results of ref.~\cite{buch} for the  neutrino mass
diagonalization in the  $(n,n)_{\mbox{\scriptsize r}}$  extension
can readily be
generalized to the  $(d,s)_{\mbox{\scriptsize r}}$  one. 
Under condition $M^N_{\mbox{\scriptsize diag}}\gg  |m|$ for all the
elements, the see-saw
neutrino mixing matrix can be found to be
\begin{equation}\label{eq:see-saw1}
{\cal U}_{m}^{n'} =
\left(
\begin{array}{cc}
1-\frac{1}{2}\xi^\dagger\xi&
\xi^\dagger\big(1-\frac{1}{2}\xi\xi^\dagger\big)\\
-\xi \big(1-\frac{1}{2}\xi^\dagger\xi\big)
&1-\frac{1}{2}\xi\xi^\dagger
\end{array}
\right)
 +{\cal O}({\xi}^4)\,,
\end{equation}
where the $s\times d$ matrix $\xi$ is 
${\xi} \equiv M^{-1} m^T$, $|\xi|\ll 1$.
Clearly, $\xi$ results in $sd$ mixing angles and $d(s-1)$ phases in
the neutrino mixing matrix.
Up to next-to-leading
order in ${\xi}$  the matrix ${\cal U}_{m}^{n'}$ brings  ${\cal
M}^n_0$
from the texture form
\begin{equation}\label{eq:see-saw3}
{\cal M}^n_0 =
\left(
\begin{array}{cc}
0&\xi^T M\\
M \xi&M 
\end{array}
\right)
\end{equation}
to the block-diagonal  form ${\cal M}^{n'}={\cal
U}_{m}^{n' T}{\cal
M}^n_0\,{\cal U}_{m}^{n'}$ with 
\begin{equation}\label{eq:see-saw2}
{\cal M}^{n'} =
\left(
\begin{array}{cc}
-\xi^T M\xi&0\\
0&M+\frac{1}{2}\Big(M \xi\xi^\dagger +\xi^* \xi^T
M\Big)
\end{array}
\right)
+{\cal O}({\xi}^3)\,.
\end{equation}
Now, by means of the unitary $d\times d$ transformation $U^{\nu'}_d$
one can diagonalize the mass matrix for light neutrinos
\begin{equation}\label{eq:constr'}
\l^{\nu}{m}^\nu_{\mbox{\scriptsize diag}}=-{U^{\nu'}_d}^{T}\xi^T\!
\l^N\! M^N_{\mbox{\scriptsize diag}}\,
\xi\, U^{\nu'}_d +{\cal O}(\xi^4)\,,
\end{equation}
so that $\xi={\cal O}((m^\nu /M^N)^{1/2})$.
Similarly, by  the unitary $s\times s$ transformation
${U^{N'}_s}=I_s+{\cal
O}(\xi^2)$ one can diagonalize the mass matrix for the heavy
neutrinos. Under condition that  the left-handed Majorana mass term
$\mu$ is ${\cal O}(1/M)$,
eq.~(\ref{eq:constr'}) straightforwardly generalizes to  
\begin{equation}\label{eq:constr'''}
\l^{\nu}{m}^\nu_{\mbox{\scriptsize diag}}={U^{\nu'}_d}^T\Big(\mu-\xi^T
\l^N M^N_{\mbox{\scriptsize diag}}
\xi\Big)\, U^{\nu'}_d
+{\cal O}(1/M^3)\,.
\end{equation}

The full neutrino mixing matrix in the see-saw
framework looks like  
${\cal U}^{n}={\cal U}_m^{n'}\,{\cal U}_d^{n'}\,{\cal U}_s^{n'}\!$.
Comparing it with that in the the doublet-singlet parametrization
eq.~(\ref{eq:U}),
one finds that the parametrizations differ by order of matrices.  As a
result, this
leads to somewhat different representations for  $V$ (and $X$). 
In the absence of the direct masses for the doublet neutrinos 
the model-independent matrix $\kappa$  is
related with the see-saw  one $\xi$ as
\begin{equation}
\xi   = (U^{\nu}_d\kappa)^\dagger  +{\cal O}(\kappa^3)\,,
\end{equation}
where $U^{\nu'}_d=U^{\nu}_d+{\cal O}(\kappa^2)$. In this, all the
quantities   $\omega$, $U^\nu_d$  and $\xi$ generally  depend on $\l$.
The parameters of
$\xi$ are  clearly shared between the independent
ones in $ U^\nu_d$ and $\omega$ in accordance with  Table~1.

In the limit 
$\,\xi\,={\cal O}(m/M)\to 0$  and hence 
${m}^\nu\to 0$,  one has to substitute effectively $U^\nu_d \to I_d$
due to the neutrino
mass degeneracy. So, all
the light neutrino mixing effects  in the
see-saw framework disappear at $v/M\ll 1$ signalling the
onset of decoupling. In particular, it follows from
eqs.~(\ref{eq:L_Y1}), (\ref{eq:L_Y2})
that Higgs boson decouples  from  the $\nu N
$ current in the see-saw framework in the leading
order  ${\cal O}(M)$, only Yukawa couplings ${\cal O}(v)$ being
generally left. As for $N N H$ vertices, they are ${\cal O}(v^2/M)$
in the limit $M\gg v$.\footnote{This contradicts the statement of 
ref.~\cite{pilaft} made in the see-saw framework on  significant
enhancement of the $\nu N H$ and $NNH$ vertices. The
enhancement could clearly take place at large $M$ only in neglect of
the  suppression ($\sim 1/M$ or $1/M^2$) of the mixing  elements.
Otherwise it could be just a numerical effect at not too large $M$.}
The see-saw
matrix $\xi$ (and more generally $\kappa$) 
results in  the non-universality and 
non-unitarity  of the lepton charged and neutral 
currents, and it can be estimated experimentally to
be small, typically $|\,\xi \,|\leq {\cal
O}(10^{-1}\div 10^{-2})$~\cite{london}.

Some comments are finally in order. 
It is clear from the above that the see-saw form of ${\cal
U}_{m}^{n'}$, given by eq.~(\ref{eq:see-saw1}),  closely resembles
the most general  one given by eq.~(\ref{eq:U_m2}).
In fact, this see-saw-like structure does not depend  on the
particular expression eq.~(\ref{eq:M_nu}) for the
neutrino mass
matrix, the latter restricting only the number of independent
parameters through constraint  eq.~(\ref{eq:constraint''}). 
Whereas the see-saw
results,  under  condition  ${m}{}^\nu\neq 0$,
can strictly  be applicable  only at  $\xi\neq 0$, the
advantage of the model-independent parametrization is that it can
straightforwardly be generalized  to a case with 
arbitrary $\kappa$. The mixings and masses become completely
disentangled. In particular, one can have, e.g.,   $\kappa=0$ at
$m{}^\nu\neq 0$, or $m{}^\nu= 0$ at
$\kappa\neq0$.
Besides, it is possible to have finite $\kappa$ at $M\gg v$ and thus
produce enhancement in the vertices with heavy
neutrinos.\footnote{Clearly, the violation of decoupling can originate
in the given framework only due to non-renormalizable
Lagrangian~(\ref{eq:DeltaL}).}
This general parametrization completely exhausts  all the
possibilities  for the neutrino masses, including these of the pure
Dirac and Majorana origins.

\section{Neutrino oscillations}

The structure of
${\cal U}^n$ for the SM general  extension $(d,s)$ (in practice,
$d=3$) could be used
when discussing the  pattern of the light neutrino oscillations.
Both Dirac and Majorana light neutrinos are permitted a priori.
Because the Dirac neutrino can be regarded as a pair of the mass
degenerate Majorana ones (with opposite CP parities), the SM
extensions with at least several
additional light degrees of freedom  are of interest. 
The primordial abundances of light nuclei in the standard big bang
nucleosynthesis restrict the effective
number of the rela\-tivistic two component interacting neutrinos 
to be $< 3.2$ ($95\%$~C.L.)~\cite{tytler}. Hence, in principle,  a
number of the light sterile neutrinos could  still be accommodated.
Sticking to as simple neutrino content  as possible 
one can encounter two different scenarios: with and without
one additional light Majorana neutrino.  

\paragraph{(i) No light singlet neutrino}

The relevant for oscillations part of the neutrino mixing matrix
${\cal U}^n$ at any $s$ reduces  in this
case to ${\cal U}^n|_{d\times d}$. In the
leading ${\cal O} (\kappa)$
approximation it is $d\times d$  unitary matrix
$U^\nu_d$. This effectively simplifies the  $(d,s)$ extension up to
$(d,0)$ in the light lepton sector (in practice, it is $(3,0)$ one
and the corresponding mixing matrix is given by eq.~(\ref{U_3})). 
As is stated before, $U^\nu_d$ depends
generally on $d(d-1)/2$ physical
mixing angles and the same number of phases. But, according to
eq.~({\ref{eq:A0}), the neutrino
oscillations  with chirality conservation (coinciding here
with the total lepton number conservation, $\Delta
L=0$)  are insensitive to $d-1$ phases capable of being resided in the
Majorana condition matrix $\l$. This reduces
the number of observable  phases to $(d-1)(d-2)/2$, exactly  as in
the Dirac case. Hence, there is  no difference here for the $\Delta
L=0$ neutrino oscillations between the 
Majorana and Dirac cases~\cite{n0}. This effective suppression  could
be evaded though for the chirality flipping (here also  lepton number
violating, $|\Delta L|=2$) oscillations. But according to
eq.~({\ref{eq:A1}) these
ones  are, in their turn, chirally suppressed, i.e., their intensity
is ${\cal O}(({m}^\nu/E)^2)$ at the neutrino energy  
$E> m^\nu$~\cite{n0,bac}. It follows that it would be hard in this
case to observe  in oscillations the Majorana specific  $CP$
violation, if any.  

Finally, in the  absence  of  light singlet
neutrinos  the chirality preserving light neutrino oscillations are
described in the given assumptions just by
the $d\times d$ unitary
matrix $U^\nu_d$ of the CKM-like type with $d(d-1)/2$ mixing
angles and $(d-1)(d-2)/2$ phases.
Account for the terms ${\cal O}(\kappa^2)$ due to the
doublet-singlet mixing  would reveal  additional $CP$ violating
phases in $U^\nu_d$ (plus those in $\kappa$ itself).
Besides, it is clear that the neutrino
oscillations in this case are mainly sensitive  to other set of the
mixing parameters than the  neutral current mixing matrix, the
latter one being determined entirely by the doublet-singlet 
mixing matrix~${\cal U}^n_m (\kappa)$. Hence in this case the two
phenomena  disentangle in essence.

\paragraph{(ii) Light singlet neutrino}

As for the case with a 
light singlet neutrino, the  doublet-singlet
mixing can no more be ignored and should be taken into
account, producing the observable effect. Among  singlet neutrinos,
only the light  one is relevant in the leading order ${\cal
O}(\kappa$) for the light neutrino mixing. One can
effectively put  in the given approximation $s=1$, thus reducing the
problem to the $(d,1)$ case.   ${\cal
U}^n_m$ in eq.~(\ref{eq:U_m2}) is given by its part 
not higher than ${\cal O}(\kappa)$,  where
$\kappa=(\kappa_1,\dots,\kappa_d)$. Thus, ${\cal U}^n$ effectively
depends on
$d(d+1)/2$ physical mixing angles and equal number of phases, in
accordance with ref.~\cite{nu1}. $d(d-1)/2$ of each
of them reside in the doublet-doublet mixing $U^\nu_d$ and $d$ in the
doublet-singlet mixing ${\cal U}^n_m(\kappa)$. Such  an approximate
$(d+1)\times (d+1)$ matrix ${\cal U}^n$ is unitary up to the given
accuracy and presents the most general  mixing matrix in this
approximation. Due to  explicit independence on $\l$ of the helicity
conserving (now do not coinciding any more with lepton number
conserving) oscillations  the number of phases relevant to these
oscillations reduces to $d(d-1)/2$, as if there were $d+1$ Dirac
neutrinos.

In reality, one has $d=3$ and experiment might
suggest a  pairwise  neutrino mixing~\cite{now}. It consists
only of the mixing  of
a pair of doublet neutrinos (chosen here as $\nu_1$ and
$\nu_2$) between themselves and the mixing of the
light singlet neutrino
($N_1\equiv\nu_4$) only with the remaining doublet
neutrino $\nu_3$, i.e., $\kappa=(0,0,\kappa_3)$.
Hence, in the case at hand the  neutrino mixing $(3,1)$ reduces to the
product of two cases $(2,0)$ and $(1,1)$, each of them corresponding
to one mixing angle and one Majorana specific phase (the latter being
unobservable in the helicity conserving oscillations). 
Under proper redefinitions, the mixing matrix ${\cal U}^n$  becomes 
\begin{equation}\label{eq:U_4}
{\cal U}^n=\left( 
\begin{array}{cc}
\left( \begin{array}{cc}
c_1&s_1\\
-s_1&c_1
\end{array}\right)
\left(\begin{array}{cc}
e^{i\gamma_{1}}&0\\
0&1
\end{array}\right)&0\\
0&\left(\begin{array}{cc}
c_{2}&s_{2}\\
-s_{2}&c_{2}
\end{array}\right)
\left(\begin{array}{cc}
e^{i\gamma_{2}}&0\\
0&1
\end{array}\right)
\end{array}
\right).
\end{equation}
In  the given assumptions, eq.~(\ref{eq:U_4}) describes  the general,
consistent with experiment, mixing for  four light neutrinos, one of
them being (quasi-)sterile. Correspondingly, one gets for the charged
current mixing matrix $V=P^{en} {\cal U}^n$
\begin{equation}
V=\left( 
\begin{array}{cccc}
c_1&s_1&0&0\\
-s_1&c_1&0&0\\
0&0&c_2&s_2
\end{array}
\right)
\left(
\begin{array}{cccc}
e^{i\gamma_1}&0&0&0\\
0&1&0&0\\
0&0&e^{i\gamma_2}&0\\
0&0&0&1
\end{array}\right)
\end{equation}
and for the neutral current one $X={\cal U}^n P^n {\cal
U}^{n\dagger}=V^\dagger V$
\begin{equation}
X=\left( 
\begin{array}{cc}
\left( \begin{array}{cc}
1&0\\
0&1
\end{array}\right)
&0\\
0&\left(\begin{array}{cc}
c_{2}^2&c_2s_{2}e^{-i\gamma_{2}}\\
c_2s_{2}e^{i\gamma_{2}}&s_2^2
\end{array}\right)
\end{array}
\right).
\end{equation}
Clearly, $c_2,s_2\neq 0$ results in  flavour violation in  neutral
currents. Remind that to completely
describe the lepton interactions
one should also specify the matrix $\eta_{CP}$ of the neutrino $CP$
signatures, as well
as  the matrix $\l$ of Majorana condition  to which
the mixing matrices above correspond. In particular, only then one can
decide whether there is $CP$ violation or not in a general case. But
in the chirality preserving oscillations, CP will always be conserved
because all the phases here are the Majorana specific ones.

\section{Summary} 

The neutrino gauge and Yukawa interactions for the SM
extensions, both renormalizable and effective, are systematically
investigated under  arbitrary Majorana condition. Independence
of the particular choice of this condition is demonstrated by means
of the explicit  rephasing invariance. The invariance is used to
exhibit manifestations of the Majorana specific phases. 
The parametrization of the neutrino mixing matrices in the
doublet-singlet factorized form is proposed. 
Its relation with the see-saw approximation is
shown. The patterns of neutrino mixing, relevant to neutrino
oscillation experiments, are exposed. 

The author is  grateful to V.V.\ Kabachenko for valuable discussions.


\begin{thebibliography}{**}

\bibitem{now}
S.M.\ Bilenky {\it et al.}, Summary of the NOW'98 Phenomenology
Working Group, hep-ph/9906251.

\bibitem{nu1}
Yu.F.\ Pirogov, Preprint IHEP 2000--2 (2000), hep-ph/0002299.

\bibitem{nm}
J.\ Schechter and J.W.F.\ Valle, Phys.\ Rev.\ {\bf D22} (1980)
2227.

\bibitem{pilaft}
A.\ Pilaftsis, Z.\ Phys.\ {\bf C55} (1992) 275, hep-ph/9901206.

\bibitem{see-saw}
T.~Yanagida, Prog.\ Theor.\ Phys.\ {\bf B135} (1978) 66; in Proc.\ of
the Workshop on Unified Theory and Baryon
Number of the Universe, eds.\ O.~Swada and A.~Sugamoto (KEK, 1979)
p.~95;
M.~Gell-Mann, P.~Ramond and R.~Slansky, in Supergravity, eds.\
P.\ van Nieuwenhuisen ans D.\ Freedman (North-Holland, Amsterdam,
1979) p.~315.

\bibitem{4f}
Yu.F.\ Pirogov and O.V.\ Zenin, Eur.\ Phys.\ J.\ {\bf C10} (1999)
629, hep-ph/9808396.

\bibitem{MC}
J.~Bernabeu and P.~Pascual, Nucl.\ Phys.\ {\bf B228} (1983) 21;
A.~Barroso and J.~Maalampi, Phys.\ Lett.\ {\bf 132B} (1983) 355; 
P.J.\  O'Donnell and U.~Sarkar, Phys.\ Rev.\ {\bf D52} (1995) 1720,
hep-ph/9305338.

\bibitem{kaiser}
B.~Kaiser, Phys.\ Rev.\  {\bf D30} (1984) 1023.

\bibitem{bil}
S.M.\ Bilenky, N.P.\ Nedelcheva and S.T.\ Petcov, Nucl.\ Phys.\ {\bf
B247} (1984) 61.

\bibitem{giunti}
C.\ Giunti, C.W.\ Kim and U.W.\ Lee,  Phys.\ Rev.\  {\bf D44} (1991)
3635, {\it ibid.}  {\bf D45} (1992) 2414; Phys.\ Lett.\ {\bf 421B}
(1998) 237, hep-ph/9709494; 
C.~Giunti, C.W.\ Kim, J.A.\ Lee and  U.W.\ Lee, Phys.\ Rev.\ {\bf
D48} (1993) 4310, hep-ph/9305276.

\bibitem{CP}
L.~Wolfenstein, Phys.\ Lett.\ {\bf 107B} (1981) 77;
J.\ Schechter and J.W.F.\ Valle, Phys.\ Rev.\ {\bf D24} (1981)
1883; Err.\ {\it ibid.} {\bf D25} (1982) 283.

\bibitem{buch}
W.~Buchm{\"u}ller and D.\ Wyler,  Phys.\ Lett.\ {\bf 249B} (1990)
458; 
W.~Buch\-m{\"u}ller and C.\ Greub, Phys.\ Lett.\ {\bf 256B} (1991)
465; Nucl.\ Phys {\bf B363} (1991) 365.

\bibitem{london}
M.\ Gronau, C.N.\ Leung and J.L.\ Rosner,  Phys.\ Rev.\ {\bf D29}
(1984) 2539; 
P.~Langacker and D.~London, Phys.\ Rev.\ {\bf D38} (1988) 886, 
907;
S.M.~Bilenky, W.\ Grimus and H.\ Neufeld, Phys.\ Lett.\ {\bf 252B}
(1990) 119.

\bibitem{tytler} 
D.~Tytler, J.M.~O'Meara, N.~Suzuki and D.~Lubin, astro-ph/0001318. 

\bibitem{n0}
J.~Schechter and J.W.F.\ Valle, Phys.\ Rev.\ {\bf D23} (1981) 1666;
M.~Doi, T.~Kotani, H.\ Nishiura, K. Okuda and E.\
Takasugi, Phys.\ Lett.\ {\bf 102B} (1981) 323.

\bibitem{bac}
J.~Bahcall and H.~Primakoff, Phys.\ Rev.\ {\bf D18} (1978) 3463.

\end{thebibliography}
\end{document}